# Direct dark mode excitation by symmetry matching of a single-particle based metasurface

Shah Nawaz BUROKUR[1,2,a)], Anatole LUPU[1,3], and André de LUSTRAC[1,2]

Affiliations:   [1]IEF, Univ. Paris-Sud, CNRS, UMR 8622, 91405 Orsay Cedex, France

[2]Univ. Paris-Ouest, 92410 Ville d'Avray, France.

[3]CNRS, Orsay, F-91405, France

Abstract: This paper makes evidence for direct dark mode excitation mechanism in a metasurface structure. The dark mode excitation mechanism is entirely determined by structures' symmetry and does not depend on near-field coupling between elements. In our examples, we consider single element based metasurface composed of two V antennas connected in an anti-symmetric arrangement. Both experimental and modeling results show an efficient excitation of magnetic dipolar mode in such structures. The direct dark mode excitation mechanism provides a design that is more robust with respect to technology imperfections. The considered approach opens promising perspectives for new type of nanostructure designs and greatly relaxes fabrication constraints for the optical domain.

[a)]Electronic Mail: shah-nawaz.burokur@u-psud.fr

1. Introduction

The concept of dark states, initially introduced in atomic physics [1-4], intensively pervaded during the last decade many of the fields of photonics such as integrated optics [5-9], photonic crystals [10,11], metamaterials [12] or plasmonics [13,14]. The interest to this subject was driven mainly by the ability to obtain an analogy of electromagnetically induced transparency (EIT). The EIT manifests as the appearance of a narrow transmission window within an absorption band [15,16]. The feature of the transmittance spectral response can be highly asymmetric. Such kind of Fano type sharp spectral features with steep intensity variation are highly desirable for sensing applications and hence generated intensive research activities on EIT at the Drude damping limit in plasmonic nanostructures and metasurfaces [12-36]. Several designs based on dolmen type geometry [13-17], plasmonic oligomers [18-23] ring-disk nanocavities [17,24,25], nanoshells [26,27], asymmetric coupled split-ring resonators [28-32] or cut wires plasmonic lattices [33-35] have been proposed and investigated for EIT characteristics.

Despite the great variety of studied designs, most of them are based on the same principle. It consists in associating a superradiant element, acting as a radiative or bright mode, with a subradiant element, playing the role of the dark (or trapped) mode. In contrast to bright mode, the dark one is only weakly coupled to free space. The resonance frequencies of the bright and dark mode are not substantially different. In a system where the constituting elements are brought far apart, the interaction between them is small. The spectral response is dominated in this case by the superradiant element. The resonance quality factor is generally low due to the strong radiation coupling. Transmission at the resonance frequency is thus highly attenuated. Making the separation distance between the elements smaller causes an increase in the near field coupling between bright and dark modes. The resulting mode hybridization leads to the opening of a narrow EIT window inside the absorption band.

With a few exceptions, the origin of the EIT in such systems was attributed to the destructive Fano interference between a directly excited bright mode with an indirectly excited dark mode. However the last theoretical advances lead to revisit this commonly shared interpretation [36-39]. In particular it was evidenced that no dark mode excitation is necessary for the existence of Fano resonances. They can be described by the interference of bright modes only. The origin of this apparent contradiction stems from the fact that the eigenmodes of the coupled system can significantly differ from those of the individual elements, and in general they are not orthogonal [39,40]. Moreover, it turns out that Fano interference effect is most pronounced when the radiative strength of the interacting modes is substantially equal. The Fano interference of two modes with substantially different radiative strength results in a very weak EIT effect [37].

In this context it is natural to wonder what is then the interest for using dark modes and if there is any other way to excite them other than through mode hybridization? The last point is especially important in view of the generally very tight fabrication tolerances for plasmonic nanostructures operating in the optical domain. The great sensitivity of the mode hybridization to the variation of the separation distance between elements on a scale of few nanometers yields highly challenging their reliable fabrication.

The aim of the present contribution is to address these critical issues. In particular we show that instead of EIT, dark mode excitation can lead also to a sharp maximum in reflection. We discuss the advantages related to this operation mode and we detail the essential of the underlying excitation mechanism which is not based on modes hybridization. We propose an excitation mechanism which is based on a direct field coupling to the dark mode by an appropriate symmetry matching with the structure geometry. The considered approach opens promising perspectives for new type of nanostructure designs and greatly relaxes technological constraints for the optical domain.

## 2. Direct dark mode excitation by symmetry matching

To present the essential of the pursued approach, let us start with the example of cut wires (CW) metasurface. We consider a uniform electromagnetic plane wave normally incident on the metasurface. For experimental facilities, demonstration of the concept is performed at microwave frequencies, keeping in mind that the extension to the optical domain can be performed in a straightforward way by structure downsizing and taking into account material properties.

To examine the behavior of the CW resonator presented in Fig. 1, the properties of the structure are calculated numerically using the Finite Element Method (FEM) Maxwell's equations solver of High Frequency Structure Simulator (HFSS) commercial code by ANSYS. The dielectric spacer used throughout this study is single face copper-cladded epoxy with a relative dielectric constant of 3.9, tangential losses of 0.02 and a thickness of 0.4 mm. For the samples reported in this work, the length of the 17 µm thick copper wires is 16.3 mm and the width is 0.3 mm.

Calculated reflection and transmission spectra of a single CW layer structure with the electric field oriented along the length of the wire are presented in Fig. 1. The transmission dip at $f_0 = 8.3$ GHz is related to the electric dipole mode excitation. It represents the lowest frequency eigenmode. Because of the invariance with respect to the charge (q → -q) and space inversion (r → -r) the eigenmodes of an isolated CW are doubly degenerated. In the presence of a capacitive interaction between adjacent cells the degeneracy is lifted. This explains the presence of two distinct dips at $f_{2s} = 17.7$ GHz and $f_{2a} = 21.2$ GHz. They come from the excitation of the second higher order mode. The lower frequency one ($f_{2s}$) corresponds to the case of a symmetric charge distribution while the higher frequency one ($f_{2a}$) to an antisymmetric charge distribution. The very sharp feature visible at 22 GHz is due to the Rayleigh anomaly caused by the opening of the first diffraction order. As evident from the spectral response, the electric dipole mode at 8.3 GHz is heavily damped. The resonance quality factor is very low $Q_0 = 1.19$ because of the strong radiative coupling to free space. In contrast, the quality factor of the second order modes is much better ($Q_{2s}$=52.2 and $Q_{2a}$=31.3) since their resonance corresponds to the higher order multipole radiation.

Besides the bright modes well visible in the spectral response, the CW structure possesses also dark eigenmodes. Their presence can be inferred from general considerations but they do not manifest in the spectral response because of the zero net dipolar momentum. So the first CW higher order eigenmode ($m_1$ in Fig. 1) that should occur in the frequency region around 15 GHz, represents a superposition of two opposite dipolar momentums $P(y) = -P(-y)$. Since the uniform electric field $E(r) = E(-r)$ is of even symmetry, the excitation of the odd symmetry doubly degenerated mode is not possible and it remains dark.

The goal of our study is to show how to excite these kind of dark modes. From symmetry considerations it follows that to avoid coupling with an uniform electric field the geometry of the resonant element should be odd [37-40]. One of the simplest odd symmetry structures is the example of two connected antisymmetric V antennas (AVA) shown in Fig. 2 [41]. The total length of the AVA in our example is set to be the same as that of the CW, *i.e.* 16.3mm. The opening angle of each V element is set to 120°. The choice of such a geometry is intentional to allow following eigenmodes evolution with respect to that of a CW. The transmission and reflection spectra at normal incidence under vertical polarization are shown in Fig. 2. When compared to the CW spectral response it can be observed that the fundamental resonance frequency $f_0 = 8.1$ GHz corresponding to the excitation of the electric dipole is very close to that of a CW. Because of the smaller capacitive coupling the splitting of the second higher order modes is greatly reduced while their mean frequency position $f_{2mean}$ = 19.7 GHz is practically the same as for the CW. Under normal incidence the modal behavior of the AVA is essentially similar to that of CW. Because of the even symmetry of the electric field and odd dark mode symmetry, the resulting interaction overlap is null and therefore the dark mode is not excited.

One solution to excite the dark mode would be to break the symmetry of the AVA structure by making for example one V antenna shorter than the other [42]. The two opposite dipolar momentums would then present different magnitudes. The resulting nonzero net dipolar moment allows a direct coupling with the electric field. The mechanism of dark mode excitation is in this case similar to the Wheatstone bridge operating principle. The coupling to the free space is due to the unbalance of electric dipolar momentums. Such solution presents the advantage of not relying on some hybridization mechanism. However the drawback is that the radiative damping is still that of an electric dipole which precludes obtaining a high resonance quality factor.

Another solution for dark mode excitation consists in using the magnetic component of the incident field. As the magnetic field transforms as a pseudovector, its symmetry is odd. The direct dark mode excitation is in this case allowed for an oblique incidence. An additional insight on the excitation mechanism can be obtained by considering the magnetic momentums generated by the loops formed by the internal angles of AVA. The dark mode excitation corresponds to the generation of currents flowing in the same clockwise or counterclockwise direction and results in a net dipolar magnetic momentum. It follows that an external magnetic field can thus directly feed the charge displacement corresponding to the dark mode excitation. It can be noted that net electric dipole momentum still remains null, contrary to the magnetic one. The transmission and reflection spectra under 45° oblique incidence for a vertically oriented electric field are shown in Fig. 2. In addition to the fundamental and second higher order modes, a new resonance appears at 14.8 GHz. Its spectral width is much narrower as compared to the fundamental resonance. Even though the dark mode excitation is achieved through a direct field coupling, the resonance quality factor is considerably increased $Q_1$=15.9. This is due to the smaller radiative damping corresponding to that of a magnetic dipole.

The fact that the dark mode manifests as a peak in reflection and not in transmission can present certain advantages. The inherent inconvenience of EIT spectral response is that it requires to create a high contrast narrow transmission band inside a high contrast reflection band. It is intuitively clear that this is more challenging as compared to create a single high contrast narrow reflection band.

It is worthwhile to note that the excitation of the dark mode resonance almost doesn't affect the rest of the spectral response. The position and intensity of the bright mode resonances associated with the fundamental and second higher order modes remain practically unchanged. On this point the dark mode excitation is essentially different from that relying on modes hybridization mechanism.

The absence of hybridization mechanism is another great advantage of the considered solution. This allows considerably relaxing the tolerances for the fabrication of nanostructures operating in the optical domain. A particular robust design can be obtained when considering an AVA design where the arms of the V elements are parallel to the borders of the unit cell. The shape of the resulting AVA looks then like a Z. The numerical modeling and experimental results for Z-shaped atom are presented in the next section.

### 3. Dark mode excitation of a Z-shaped meta-atom: simulations and experiments

The geometry of the Z shaped is represented in Fig. 3(a). It is composed of two AVA with V antenna internal angle of 45°. As for the previous case of 120° AVA the total length of the Z is set the same as that of the CW (16.3 mm). Fundamentally, the dark mode excitation mechanism is the same as for AVA. The essential difference of the Z atom with respect to the 120° AVA is that Z legs parallel to the borders of unit cell are bringing some advantages.

It can be observed that Z legs can act as an additional capacitance and also as a current loop. As represented in Fig. 3(a) for the fundamental resonance, the flow of the current leads to the accumulation of the opposite sign charges between the adjacent Z legs. This creates an additional capacitance that causes the fundamental resonance to shift to lower frequencies. At the same time, since the direction of the current flow (indicated by the red arrows) is the same for the adjacent Z legs, no dipolar magnetic momentum is created. The fundamental resonance is essentially that of an electric dipole.

For the first higher order resonance, which is that of a dark mode, due to the opposite direction of the currents flow in the adjacent Z legs, the accumulated charges are of the same sign. This lowers the capacitance of the structure and shifts up the resonance frequency. At the same time the oppositely directed currents flow in the adjacent Z legs create an additional loop making the dark mode excitation more efficient. The numerical modeling and experimental results for Z atom shown in Figs. 3(b)-(d) confirm the formulated predictions.

Calculated reflection and transmission spectra of a single Z layer structure with the electric field polarized vertically show a resonance around 6 GHz for all incidence angles, that corresponds to the lowest LC frequency of the Z-shaped meta-atom. Also for off-normal incidence in the H-plane (plane containing vectors H and k) with the electric field still vertically aligned to the Z meta-atom (left column in Figs. 3(b)-(d)), the excitation of an additional resonance can be observed around 17.9 GHz. The excitation of the dark mode in this case results in the appearance of a narrowband resonance in reflection. As it can observed in the reflection responses under H-plane incidence, the magnitude of the resonance varies from 0.41 for 15° to 0.86 for 45°. We note the very good qualitative and quantitative agreement between modeling predictions and experimental results.

The excitation of the dark mode is highly efficient under the H-plane oblique incidence and totally negligible under the E-plane incidence (right column in Figs. 3(b)-(d)), *i.e.* when there is no magnetic field component perpendicular to the plane of Z atom. The measured experimental results thus prove the direct dark mode excitation mechanism by the magnetic component of the incident field.

To demonstrate that dark mode excitation is not relying on any hybridization mechanism, we perform modeling of a structure where dimensions of the Z atom are the same but separation distance between adjacent Z elements is increased from 0.2 mm to 1.4 mm. As evident from results displayed in Fig. 4, the efficiency of dark mode excitation for off-normal incidence in the H-plane is very little affected, despite the important change in coupling between the adjacent elements. The important change of the coupling strength can be directly inferred from the observed variation of the resonance frequencies.

By changing the lattice period to $p + 20\%$ with $p = 6$ mm, the capacitance between the adjacent Z legs decreases making the electric dipolar resonance frequency shift to 7.9 GHz, as shown in Fig. 4. For the same reason, but making this time an opposite effect, the dark mode excitation under oblique incidence appears at a slightly lower frequency (16.4 GHz) than in the original case with lattice period $p$ (Fig. 3). The results presents in Fig. 4 justifies the fact that the dark mode excitation does absolutely not rely on any modes hybridization, but on a direct field coupling through an appropriate symmetry matching with the structure geometry.

When the Z meta-atom is rotated by 22.5° around the *z*-axis (propagation direction), and keeping the vertical polarization configuration, the dark mode excitation is forbidden under normal incidence. Conversely, as shown in Fig. 5, dark mode excitation is very efficient under oblique incidence when there is a magnetic field component perpendicular to the incidence plane.

All provided examples show that for dark mode excitation relies only on symmetry matching and doesn't depend of coupling between elements. The direct dark mode excitation mechanism provide thus a design that is more robust with respect to technology imperfections and greatly relaxes fabrication tolerances. This moment is of great importance when considering fabrication of nanostructures operating in the optical domain [43].

4. **Summary and conclusions**

The present contribution address the problematic of dark mode excitation in metasurface structures different from that using mode hybridization mechanism. We propose a dark mode excitation mechanism based only on symmetry matching conditions. It is shown that direct excitation mechanism is possible for an anti-symmetric type higher order mode having a zero net electric dipolar momentum but different from zero magnetic one. The excitation of magnetic dipolar momentum can be achieved under field oblique incidence on metasurface having anti-symmetric unit cell geometry. In our examples we considered single-element metasurface composed of two V antennas connected in an anti-symmetric arrangement or more simply Z-shaped meta-atoms. Both experimental and modeling results show an efficient excitation of magnetic dipolar mode in such structures.

The great advantage of the considered approach is that dark mode excitation is entirely determined by structures' geometry symmetry and doesn't depend of coupling between elements. The considered approach opens promising perspectives for new type of nanostructure designs and greatly relaxes technological constraints for the optical domain.

**Figure captions**

Figure 1: Cut-wire array under normal incidence. The length and width of the copper wires are respectively 16.3 mm and 0.3 mm. (a) Reflection and transmission characteristics under normal incidence. (b) The charge distribution for the first three resonances are shown. Dark mode ($m_1$) is not excited under normal incidence due to zero net dipolar momentum.

Figure 2: Reflection and transmission characteristics of an array of two connected antisymmetric V antennas. In contrast to normal incidence, dark mode is excited under 45° oblique incidence, due to a non-zero net dipolar magnetic momentum.

Figure 3: (a) Charge distribution at fundamental and first higher order resonances. (b)-(c) Calculated reflection and transmission. (d) Measured transmittance.

Figure 4: Evidence of dark mode excitation by direct field coupling. The separation between adjacent Z atoms is increased, therefore introducing less inter-cell capacitance.

Figure 5: Reflection and transmission characteristics of a twisted Z atom. (a)-(b) Calculations. (c) Measurements.

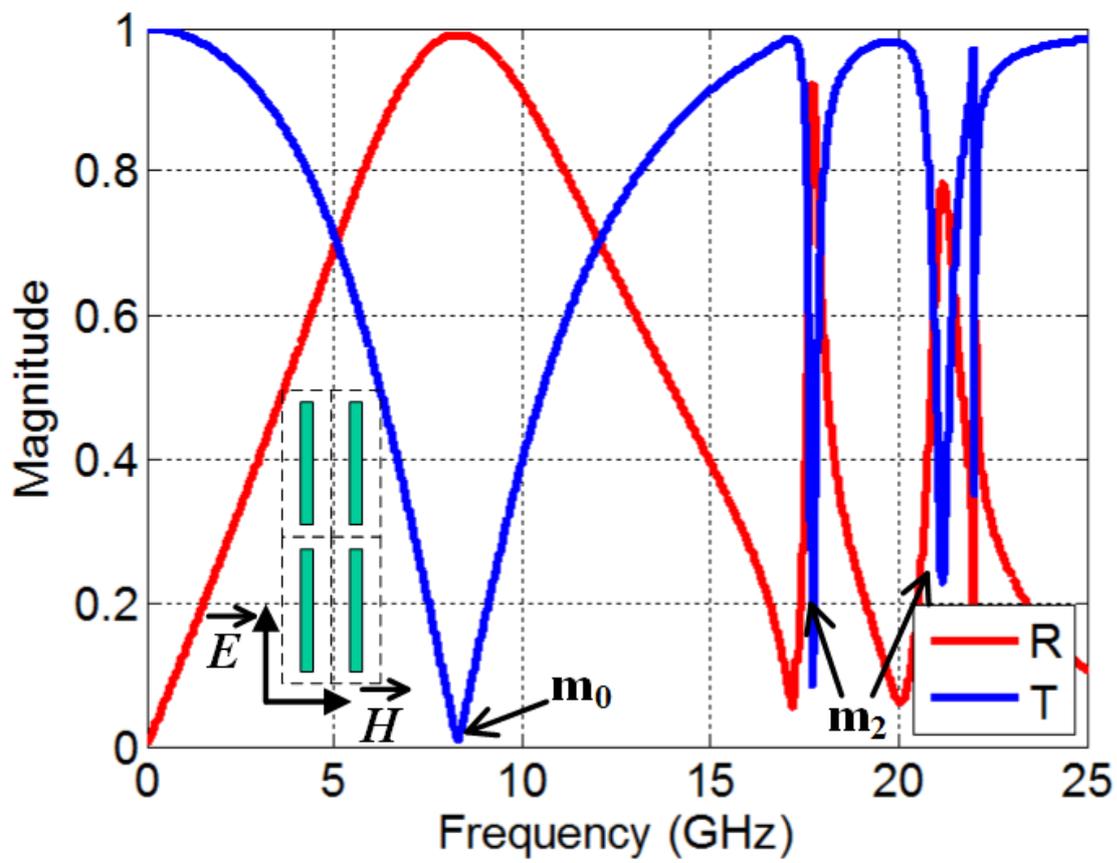

(a)

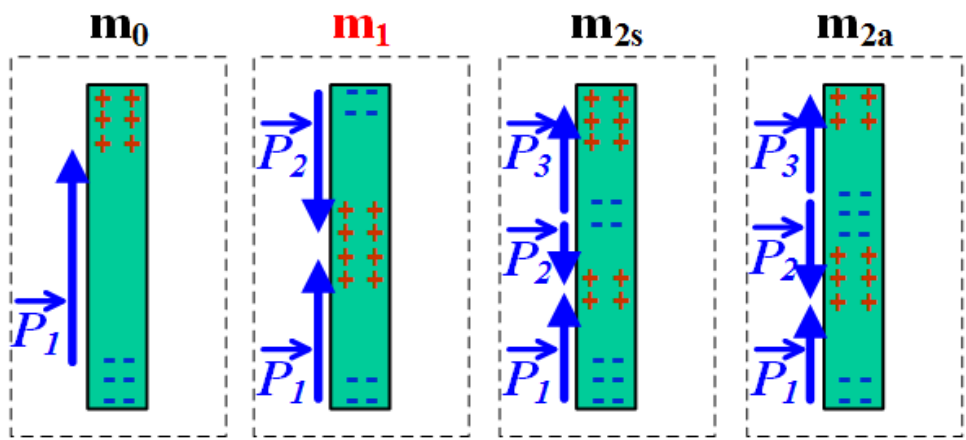

(b)

Fig. 1

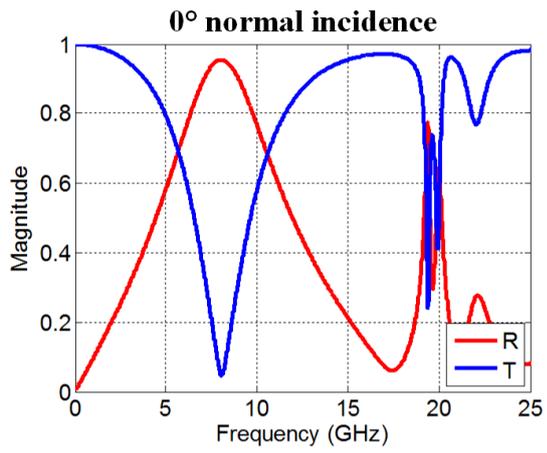
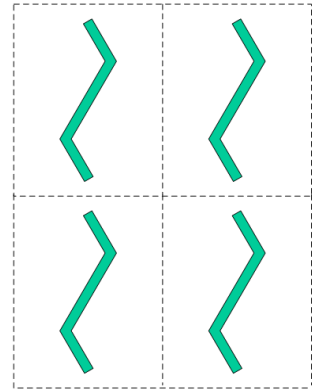
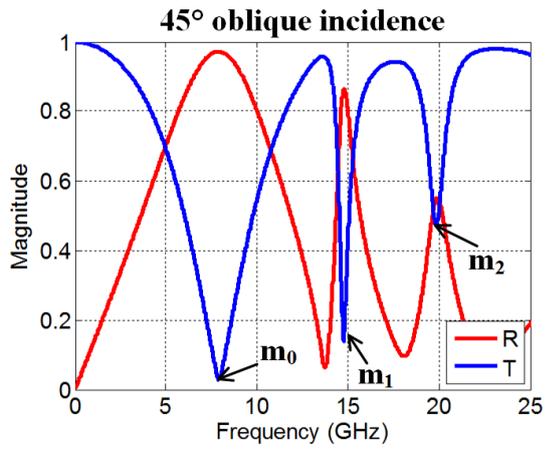
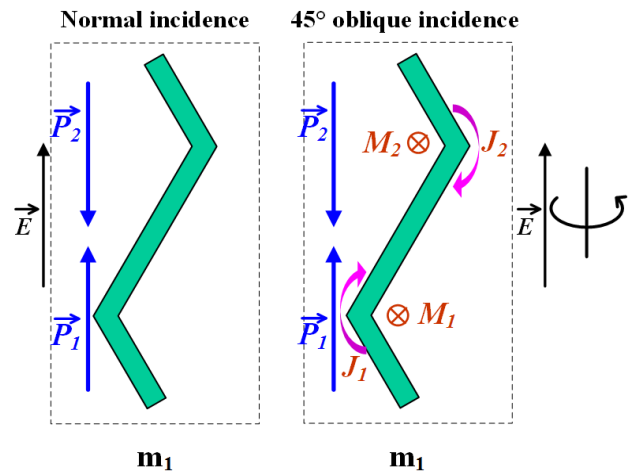

Fig. 2

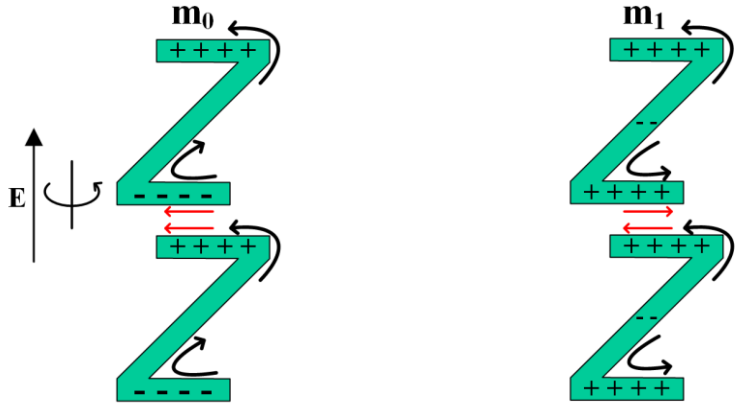

(a)

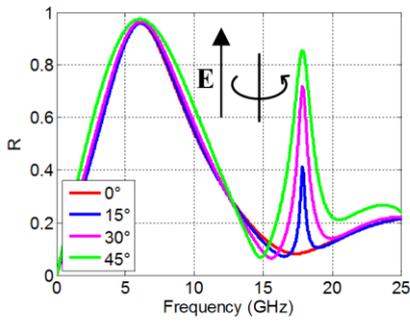 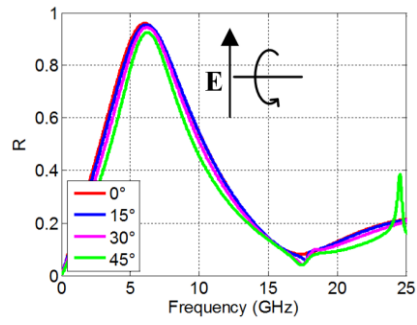

(b)

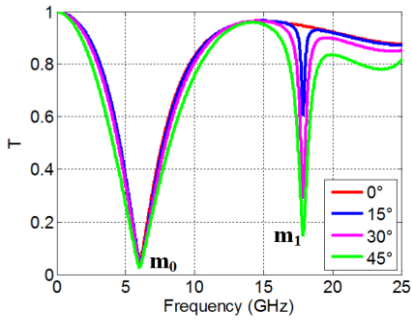 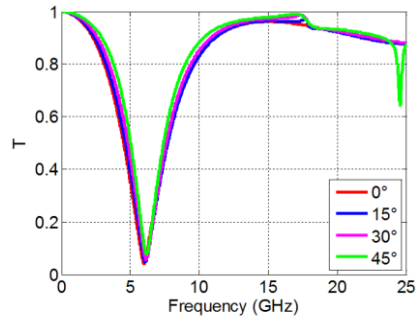

(c)

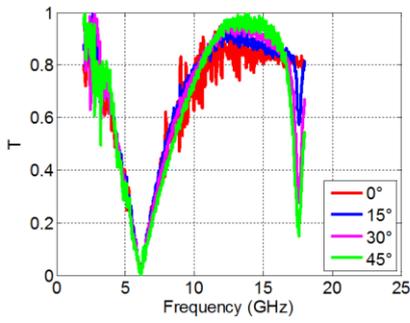 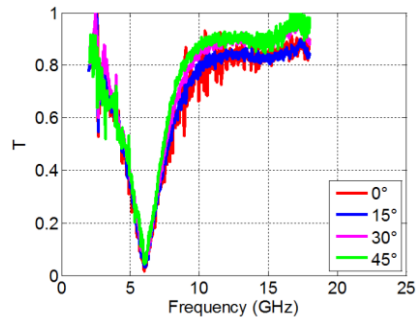

(d)

Fig. 3

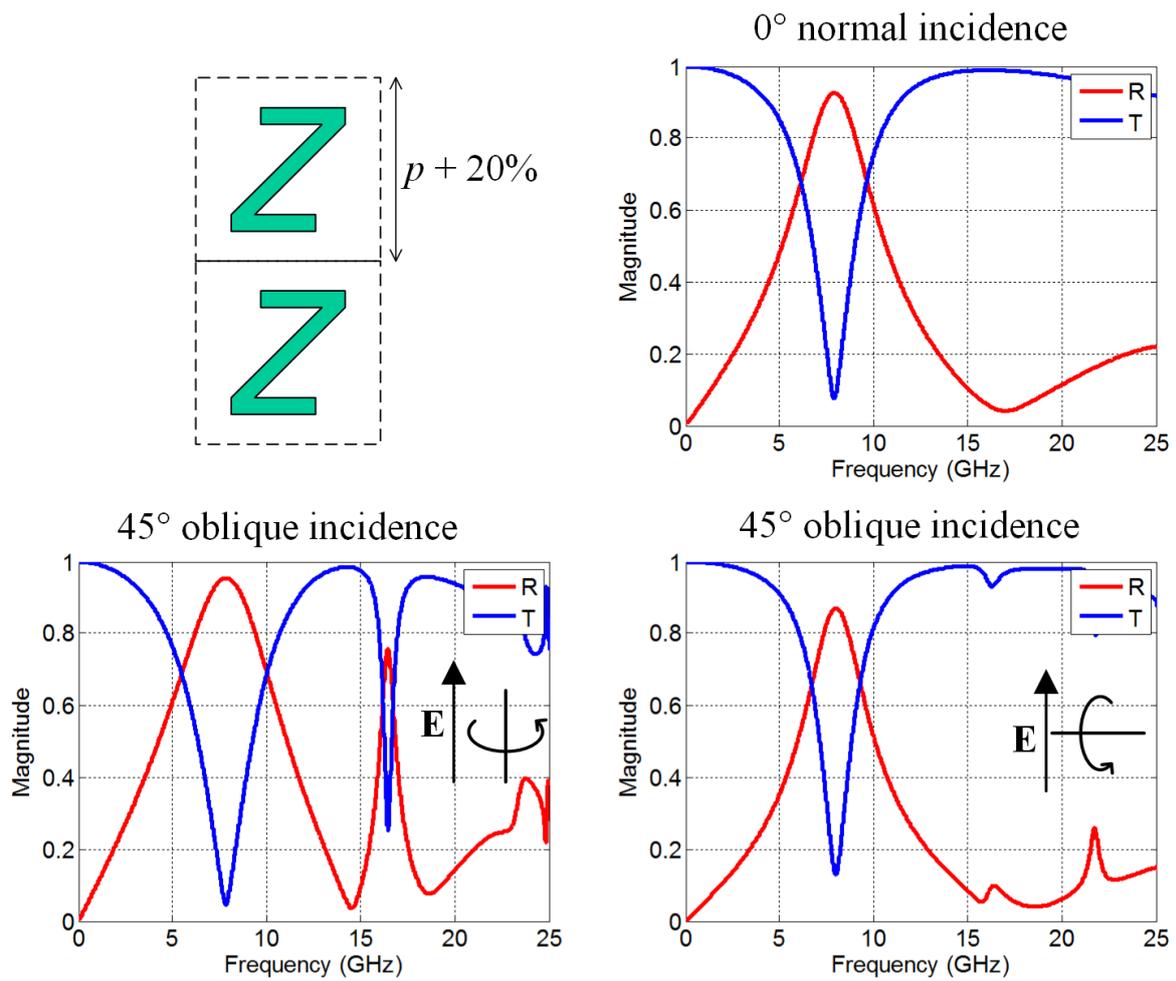

Fig. 4

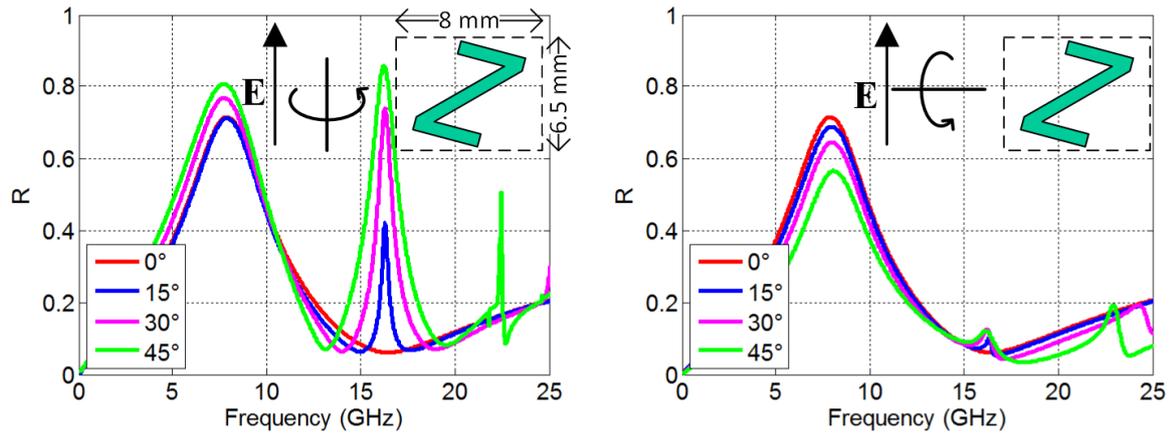

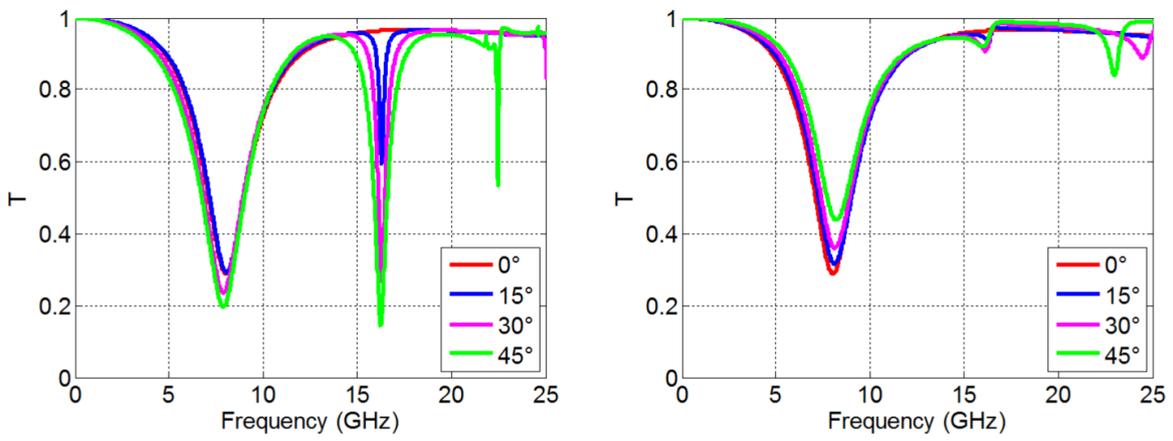

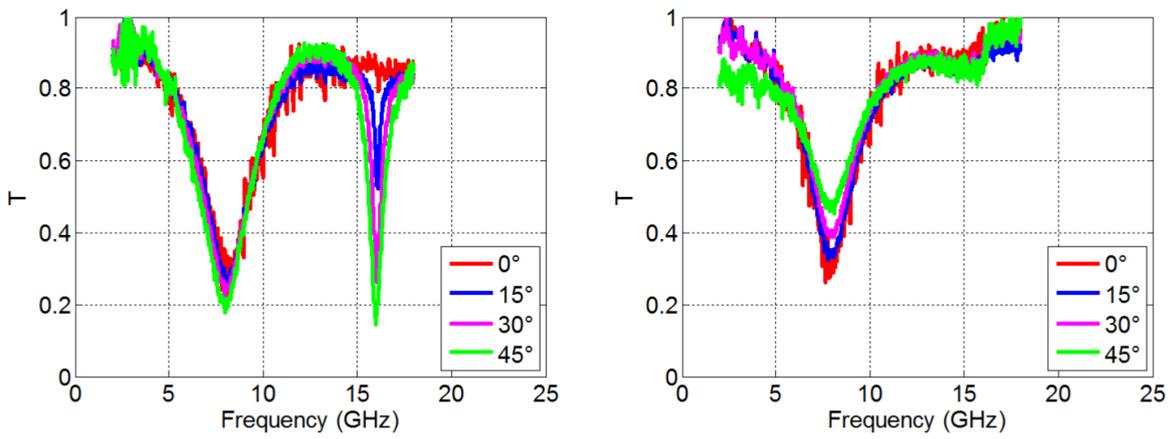

Fig. 5